\documentclass[aps,prb,twocolumn,showpacs,preprintnumbers,amsmath,amssymb]{revtex4}

\usepackage{graphicx} 
\usepackage{dcolumn} 
\usepackage{bm} 
\usepackage{epsfig}
\usepackage{amssymb}

\begin{document}
\newcommand{\kp}{$\mathbf{k}\!\cdot\!\mathbf{p}$}
\def\e{\mathop{\rm \mbox{{\Large e}}}\nolimits}
\def\im{\mathop{\rm \od{\iota}}\nolimits}
\newcommand{\ts}[1]{\textstyle #1}
\newcommand{\bn}[1]{\mbox{\boldmath $#1$}}
\newcommand{\bc}{\begin{center}}
\newcommand{\ec}{\end{center}}
\newcommand{\be}{\begin{equation}}
\newcommand{\ee}{\end{equation}}
\newcommand{\bea}{\begin{eqnarray}}
\newcommand{\eea}{\end{eqnarray}}
\newcommand{\ba}{\begin{array}}
\newcommand{\ea}{\end{array}}

\title{Plasmon polaritons in photonic superlattices containing a left-handed material}

\author{E. Reyes-G\'omez$^{1}$, A. Bruno-Alfonso$^{2}$, S. B.
Cavalcanti$^{3,4}$, C. A. A. de Carvalho$^{4,5}$, and L. E.
Oliveira$^{4,6}$ } \affiliation{$^{1}$Instituto de F\'{\i}sica,
Universidad de Antioquia, AA 1226, Medell\'{\i}n, Colombia \\
$^2$Faculdade de Ci\^encias, UNESP - Universidade Estadual
Paulista, 17033-360, Bauru-SP, Brazil \\
$^3$Instituto de
F\'{\i}sica, UFAL, Cidade Universit\'{a}ria,
57072-970, Macei\'{o}-AL, Brazil \\
$^4$Inmetro, Campus de Xer\'{e}m, Duque de Caxias-RJ, 25250-020, Brazil \\
$^5$Instituto de F\'{\i}sica, UFRJ, Rio de Janeiro-RJ, 21945-972, Brazil \\
$^6$Instituto de F\'{\i}sica, UNICAMP, CP 6165, Campinas-SP,
13083-970, Brazil}

\date{\today}

\begin{abstract}

We analyze one-dimensional photonic superlattices which
alternate layers of air and a left-handed material. We assume
Drude-type dispersive responses for the dielectric permittivity and magnetic
permeability of the left-handed material. Maxwell's equations and
the transfer-matrix technique are used to derive
the dispersion relation for the propagation of obliquely incident
optical fields. The photonic dispersion indicates that the
growth-direction component of the electric (or magnetic) field leads to the propagation of
electric (or magnetic) plasmon polaritons, for either TE
or TM configurations. Furthermore, we show that if the plasma
frequency is chosen within the photonic $\langle n(\omega)\rangle=0$
zeroth-order bandgap, the coupling of light with plasmons
weakens considerably. As light propagation is forbidden in that
particular frequency region, the plasmon-polariton
mode reduces to a pure plasmon mode.

\end{abstract}


\maketitle

Over the years, artificial complex materials have been
increasingly used to shape and manipulate light
\cite{Shelby01,Barnes03,Maier05,Rama05,Ozbay06}. The microstructuring
of high quality optical materials yields remarkable
flexibility in the fabrication of nanostructures, and
allows for the tailoring of electromagnetic dispersions
and mode structures to suit almost any need. Metamaterials
\cite{Dolling,Zhu,Dragoman,Pacheco2002,Elef2002,Grbic2002,Li2003,Jiang2003}, also known
as left-handed materials (LHMs), are a remarkable example of such
nanostructuring. Light propagation through metamaterials is
characterized by a phase velocity opposite to the Poyinting
vector, which corresponds to negative dispersive electric and
magnetic responses.

The advent of metamaterials has opened a new era for optical
devices, and has also given considerable thrust to the recent area
of plasmonics, the study of plasmon polaritons. In metal-dielectric
interfaces, for example, surface-plasmon polaritons  are coupled modes
that result from resonant interactions between
electromagnetic waves and mobile electrons at the surface of
a metal or semiconductor. Such resonant surface-plasmon polaritons
may have much shorter wavelengths than that of the radiation, which enables them
to propagate along nanoscale systems
\cite{Shelby01,Barnes03,Maier05,Rama05}, opening up a wide
range of possibilities for the construction of new optical
devices. It is well known, for example, that one of the most important
features of surface-plasmon polaritons is to
confine light to very small dimensions,
yielding the merging of photonics and electronics at the nanoscale. Furthermore, considering the dispersive character of the LHM's,
together with the enhanced optical magnetism exhibited by them, one might conjecture the existence
of remarkable new phenomena such as the excitation of plasmon polaritons of a
magnetic nature, that is, magnetic density waves resulting from resonant
interactions between the optical field and current densities at the
metamaterial.  Ultimately, those developments could lead to increases in the resolution
of microscopes, in the efficiency of LEDs, and in the sensitivity
of chemical and biological devices \cite{Ozbay06,Dolling,Zhu,Dragoman}.

\begin{figure}
\epsfig{file=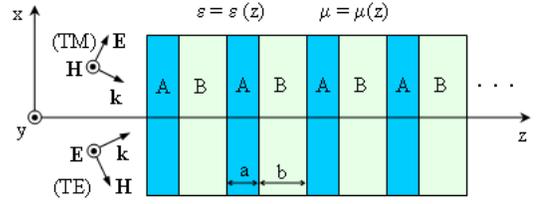,width=0.8\columnwidth} %
\caption{(Color online) Pictorial view of the 1D multilayer
photonic superlattice with layers A and B in periodic arrangement,
and the electric and magnetic fields for the TE-like and TM-like
electromagnetic waves schematically shown. Note that, for normal
incidence, the two polarizations are equivalent.}\label{fig1}
\end{figure}

One-dimensional (1D) superlattices which alternate layers of
positive and negative materials have already exhibited many interesting
properties
\cite{Smith00,Lisci06,Cavalcanti,Shadrivov03,Zio04,Daninthe06,Yuan06,Euden07,Weng07,Zhang07}
that are absent in superlattices composed solely of positive materials. In
particular, the existence of a non-Bragg photonic bandgap, also
known as a zeroth order gap, has been suggested \cite{Li2003},
detected \cite{Jiang2003,Smith00}, and characterized
\cite{Lisci06,Cavalcanti}. In order to investigate the possibility of excitation of electric/magnetic plasmon polaritons, 
in the present work we study the oblique incidence of light on a model 1D superlattice composed of layers A of air, and layers B of a doubly
negative material.  Layers A (width $a$) and B (width $b$) are
distributed  periodically so that $d=a+b$ is the period of the
superlattice nanostructure (cf. Fig. \ref{fig1}).

In the B
layers, the electric and magnetic responses are dispersive and may
assume negative values. If one neglects losses, they may be
described by
\cite{Pacheco2002,Elef2002,Grbic2002,Li2003,Jiang2003}
\begin{equation}  \label{e01}
\varepsilon_B(\omega)=\varepsilon_0-\frac{\omega_{e}^{2}}{\omega^2}\,\,\,;\,\,\,
\mu_B(\omega)=\mu_0-\frac{\omega_{m}^{2}}{\omega^2} \,,
\end{equation}
where $\varepsilon_B(\omega)$ and $\mu_B(\omega)$ are the
dielectric permittivity and magnetic permeability in slab B, respectively, one may choose \cite{Elef2002,Grbic2002} $\varepsilon_0 = 1.21$ and $\mu_0 = 1.0$,
and the electric/magnetic plasmon
modes are at $\nu=\nu_{e} = \frac{\omega_e} {2 \pi \sqrt{\epsilon_0}}$
and $\nu=\nu_{m} = \frac{\omega_m} {2 \pi \sqrt{\mu_0}}$, which correspond to the
solutions of $\epsilon_B(\omega)=0$ and $\mu_B(\omega)=0$, respectively.

We note that dispersions such as those in \eqref{e01} hold in periodically LC
loaded transmission lines \cite{Elef2002}. These negative-index
systems were shown to exhibit good microwave properties, with low
loss and broad bandwidth \cite{Grbic2002}.

We shall be
interested in studying the properties of both the
transverse-electric (TE: electric field parallel to
the interface plane, see Fig. 1) and transverse-magnetic (TM:
magnetic field parallel to the interface) polarizations of a
monochromatic electromagnetic field of frequency $\omega$
propagating through a 1D periodic system.

In the case of a TE field, one may choose
\begin{equation}
\label{e09}
\mathbf{E} (\mathbf{r},t) = E (z) \exp \left [ i \left ( q x - \omega t \right ) \right ] \mathbf{e}_y \,,
\end{equation}
whereas, in the case of TM polarization, the magnetic field may be
considered as
\begin{equation}
\label{e10}
\mathbf{H} (\mathbf{r},t) = H (z) \exp \left [ i \left ( q x - \omega t \right ) \right ] \mathbf{e}_y,
\end{equation}
where we have assumed that the superlattice was grown along the $z$
axis, $q$ is the wavevector component along the $x$ direction,
and $\mathbf{e}_y$ is the cartesian unitary vector along the $y$
direction.

Maxwell's equations lead to the following differential equations for the electric and magnetic amplitudes
\begin{equation}
\label{e11}
\frac{d}{dz} \left[\frac{1}{\mu(z)} \frac{d}{dz} E(z)\right] = - \epsilon (z) \left [ \left ( \frac{\omega}{c} \right )^2 - \frac{q^2}{n^2 (z)}  \right ] E (z),
\end{equation}
and
\begin{equation}
\label{e12}
\frac{d}{dz} \left[\frac{1}{\epsilon(z)} \frac{d}{dz} H(z)\right] = - \mu (z) \left [ \left ( \frac{\omega}{c} \right )^2 - \frac{q^2}{n^2 (z)}  \right ] H (z),
\end{equation}
where $n (z) = \sqrt {\epsilon (z)} \sqrt{\mu (z)}$ is the $z$-position dependent refraction index.

In the sequel, $n_A$ and $n_B$ are the refraction indices, while
$\mu_A$ and $\mu_B$ are the magnetic permeabilities in layers A
and B, respectively. Note that, because of (\ref{e01}), the $n_B$
refraction index in layer B is a function of the frequency, and
may be a real positive, real negative or a pure imaginary number.
Also, note that $q$, the incident wavevector component along the
$x$ direction, may be obtained as a function of the angle of
incidence $\theta\equiv\theta_A$ by $q = \frac{\omega}{c} n_A  \sin\theta$.

Equations (\ref{e11}) and (\ref{e12}) may be solved by means of
the transfer-matrix technique (see, for example, the work by
Cavalcanti \textit{et al} \cite{Cavalcanti} and references
therein). For $n_B$ a real number, and $n_B^2 - n_A^2 \sin^2
\theta \geq 0$, the procedure yields the TE polarization
dispersion relation from the solution of the transcendental
equation
\begin{eqnarray}
\label{e13}
\cos(k d) &=&  \cos (Q_A a) \cos (Q_B b) \nonumber \\
          &&- {\frac{1}{2}\left(\frac{F_A}{F_B}+ \frac{F_B}{F_A}\right)} \sin (Q_A a) \sin (Q_B b).
\end{eqnarray}
In the formula, $k$ is the Bloch wavevector along the $z$
direction which is the axis of the photonic crystal; fields in
consecutive unit cells are related by the Bloch condition, i.e.,
by the phase factor $e^{\,i\,k\,d}$. $Q_A$ and $Q_B$ are defined
as
\begin{equation}
\label{e15}
Q_A = \frac{\omega}{c} n_A  \vert \cos \theta \vert=\frac{\omega}{c} n_A  \vert \cos \theta_A \vert,
\end{equation}
and
\begin{equation}
\label{e16}
Q_B = \frac{\omega}{c} \sqrt{n_B^2 - n_A^2 \sin^2 \theta}= \frac{\omega}{c} |n_B|  \vert \cos \theta_B \vert,
\end{equation}
where, $F_A=(Q_A/\mu_A)$, $F_B=(Q_B/\mu_B)$ and, in the last
equality, we have made use of Snell's law $n_A\sin \theta=n_B\sin
\theta_B$.

\begin{figure}
\epsfig{file=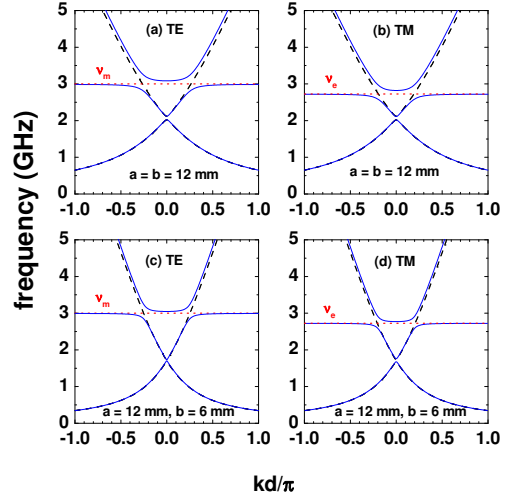,width=0.75\columnwidth} %
\caption{(Color online) TE and TM dispersion relations $\nu = \nu
(k)$ in photonic periodic superlattices ($\nu = \frac{\omega}{2\pi}$). Calculations were
performed for air as slab A ($\epsilon_A = 1$, $\mu_A =1$), $a = b = 12$ mm (also
for $a = 12$ mm and $b = 6$ mm), and $\omega_e /2 \pi = \omega_m /2 \pi = 3$ GHz  for
the Drude model [cf. Eq. (\ref {e01})] in slab B. Dotted lines indicate the pure electric/magnetic plasmon modes
at $\nu=\nu_{e} = \frac{\omega_e} {2 \pi \sqrt{\epsilon_0}}$ and $\nu=\nu_{m} = \frac{\omega_m} {2 \pi \sqrt{\mu_0}}$. Solid and dashed lines correspond to $\theta = \pi/12$ and $\theta = 0$ incidence angles, respectively.}\label{fig2}
\end{figure}

\begin{figure}
\epsfig{file=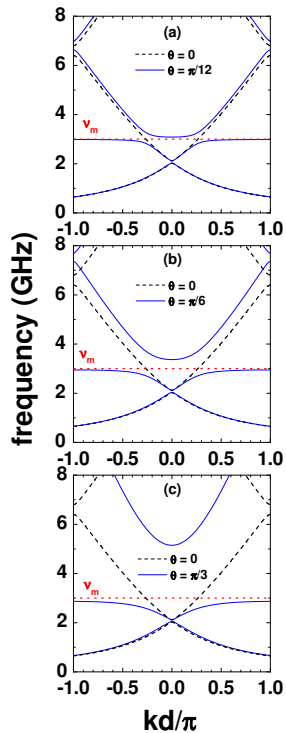,width=.42\columnwidth} %
\caption{(Color online) TE dispersion relations $\nu = \nu
(k)$ in photonic periodic superlattices ($\nu = \frac{\omega}{2\pi}$) as functions of various
incidence angles. Calculations were
performed using the same parameters as in Fig. 2. Dotted lines indicate the pure magnetic
plasmon modes
at $\nu=\nu_{m} = \frac{\omega_m} {2 \pi \sqrt{\mu_0}}$, dashed lines correspond to $\theta = 0$, and
solid lines correspond to (a) $\theta = \pi/12$, (b) $\theta = \pi/6$ and (c) $\theta = \pi/3$
incidence angles, respectively.}\label{fig3}
\end{figure}

\begin{figure}
\epsfig{file=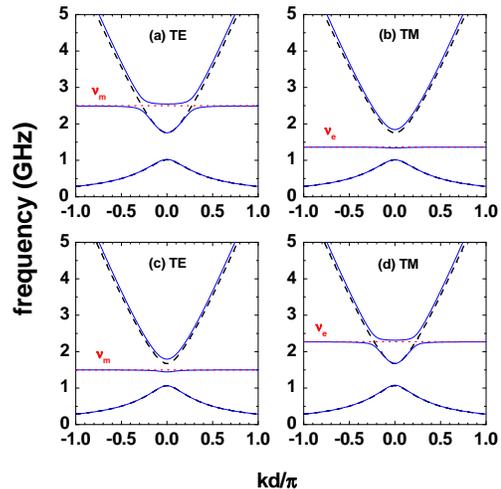,width=0.75\columnwidth} %
\caption{(Color online) TE and TM dispersions in a
photonic periodic superlattice with $a = b =$ 12 mm. Numerical
calculations were carried out for air as slab A ($\epsilon_A = 1$ and $\mu_A =1$).
For slab B, $\epsilon_B$ and $\mu_B$ were chosen by taking plasma frequencies in the Drude
model as $\omega_e /2 \pi = 1.5$ GHz and $\omega_m /2 \pi = 2.5$ GHz,
respectively, in (a) and (b), whereas in (c) and (d) we have taken
$\omega_e /2 \pi = 2.5$ GHz and $\omega_m /2 \pi = 1.5$ GHz. Note that, in panel (b),
the electric plasmon frequency $\nu_{e} = \frac{\omega_e} {2 \pi \sqrt{\epsilon_0}}$ is in
the $\langle n(\omega)\rangle=0$ zeroth-order bandgap region, leading to a basically flat
$\theta = \pi/12$ plasmon-polariton band with a frequency essentially equal to
$\nu_e$ for the TM dispersion. A similar situation is observed
for the TE dispersion in (c), as $\nu_{m} = \frac{\omega_m} {2 \pi \sqrt{\mu_0}}$ falls in
the zeroth-order gap region.
Solid and dashed lines correspond to $\theta = \pi/12$ and $\theta = 0$ incidence angles,
respectively.}\label{fig4}
\end{figure}

For $n_B$ a real number, and $n_B^2 - n_A^2 \sin^2 \theta < 0$,
the procedure yields the TE polarization dispersion relation from
the solution of the transcendental equation
\begin{eqnarray}
\label{e17}
\cos(k d) &=&  \cos (Q_A a) \cosh (Q_B b) \nonumber \\
          &&- {\frac{1}{2}\left(\frac{F_A}{F_B}- \frac{F_B}{F_A}\right)} \sin (Q_A a) \sinh (Q_B b).
\end{eqnarray}
where $Q_A$ is still given by (\ref{e15}), but
\begin{equation}
\label{e18}
Q_B = \frac{\omega}{c} \sqrt{n_A^2 \sin^2 \theta - n_B^2}.
\end{equation}
Moreover, if $n_B^2 < 0$, (\ref{e17}) is still valid, with $Q_A$ still given by (\ref{e15}), but
\begin{equation}
\label{e19}
Q_B = \frac{\omega}{c} \sqrt{n_{BI}^2 - n_A^2 \sin^2 \theta},
\end{equation}
where $n_{BI}$ is the imaginary part of $n_B$.

For the TM polarization, the transcendental equations (\ref{e13})
and (\ref{e17}), as well as the definitions of $Q_A$ and $Q_B$ for
the  different cases are also valid, provided one replaces
$\mu_A$ by $\epsilon_A$ and $\mu_B$ by $\epsilon_B$, where
$\epsilon_A$ and $\epsilon_B$ are the dielectric permissivities in
layers A and B, respectively.

Figure~\ref{fig2} shows the calculated dispersions ($\nu = \frac{\omega}{2\pi}$) in the cases of TE
[Figs.~\ref{fig2}(a) and~\ref{fig2}(c)] and TM
[Figs.~\ref{fig2}(b) and~\ref{fig2}(d)] polarizations, for
different layer widths. Note that $\nu_{e}$ and $\nu_{m}$ fall outside the $\langle
n(\omega)\rangle=0$ zeroth order bandgap. For $\theta =
0$ (dashed curves), the second photonic band, just above the
$\langle n(\omega)\rangle=0$ zeroth order gap, is a pure
photonic mode and, for $\theta = \pi/12$, there is a coupling
between the radiation field and a plasmon mode which
leads to a pair of plasmon-polariton modes. In other words, while for
normal incidence the band just above the zeroth order gap is a
pure photonic mode, for oblique incidence ($\theta \neq 0$) the \emph{electromagnetic field +
plasmon} interaction leads to a pair of coupled modes. We emphasize that, for $\theta =
0$, the electric- and magnetic-field components along the $z$ growth direction are null and there is no
longitudinal wave propagation. On the other hand, for $\theta \neq 0$, in the case of the TM (TE) configuration, the
electric (magnetic) field has a component on the $z$ direction, which results in the excitation of longitudinal electric (magnetic)
waves. For $\theta = \pi/12$, Figure~\ref{fig2} illustrates that the pair of plasmon-polariton bands are asymptotic to
the $\nu = \nu_{e}$ and $\nu = \nu_{m}$ pure
electric/magnetic plasmon values. It is clear from
Fig.~\ref{fig2} (a), for example, that at small values of $k$, the lowest
$\theta = \pi/12$ plasmon-polariton mode behaves like an electromagnetic
photonic wave. As $k$ increases, the dispersion
bends over and reaches the limiting value of the pure magnetic plasmon. On the other hand, the second $\theta =
\pi/12$ plasmon-polariton branch behaves like a magnetic plasmon wave, at
small $k$, and like an electromagnetic photonic mode as $k$ increases.
The same comments are valid for the other panels in
Fig.~\ref{fig2} [of course, in the cases of Figs.~\ref{fig2}(b)
and~\ref{fig2}(d), the electric plasmon mode is the one
involved]. By comparing the results for normal and oblique
incidence, it is clear that, for $\theta \neq 0$, resonant plasmon-polariton
waves are excited by the coupling of the electric (or magnetic)
plasmon modes with the incident eletromagnetic field.
One may infer that the plasmon-polariton waves are excited by the magnetic field, in the TE case, or
driven by the electric field, in the TM case. This is
consistent with the fact that those longitudinal waves,
in the TE case, are asymptotic to the $\nu_{m}$ pure magnetic
plasmon frequency whereas, in the TM case, the plasmon-polariton bands are asymptotic to
the $\nu_{e}$ pure electric plasmon mode.

The calculated
dispersions, for TE
modes, as a function of the incidence angles are displayed in Fig.~\ref{fig3}. We notice that the essential features persist
irrespective of the incidence angle, although the Bragg gaps widens considerably for higher
incidence angles. Also, Fig.~\ref{fig4} clearly indicates that, by choosing the
resonant plasmon frequency within the $\langle
n(\omega)\rangle=0$ zeroth order bandgap, the coupling of light to
plasmon modes, for $\theta \neq 0$, essentially disappears leading to a
basically pure (electric or magnetic) plasmon mode. This
is a consequence of the fact that the energy of the incident
electromagnetic wave lies in a forbidden energy gap region and,
therefore, the coupling of the incident light with
plasmons is expected to be quite weak.

Summing up, we have studied light propagation through a 1D
superlattice composed of alternate layers of a positive constant
material and a negative dispersive material, and verified the
appearance of coupled plasmon-polariton modes of electric and magnetic natures.
The photonic dispersion was
calculated from Maxwell's equations using the
transfer-matrix approach, with the LHM modeled by Drude-like
dielectric and magnetic responses. In the case of oblique $\theta
\neq 0$ incidence, the photonic dispersion indicates that the magnetic or electric field in the
frequency region around the $\langle n(\omega)\rangle=0$ zeroth
order bandgap leads to coupled magnetic or electric plasmon-polariton modes for the TE and TM configurations, respectively. Moreover,
present results show that the coupling of light with plasmons
is weakened by choosing the plasma frequency
inside the zeroth order gap. As light propagation is forbidden in
that particular gap-frequency region, the coupled plasmon-polariton mode becomes
essentially a pure plasmon mode. This feature permits one to select which type of plasmon polariton
(i.e., electric or magnetic)
one is willing to excite, by choosing the magnetic or electric plasmon frequency within the
zeroth order gap. Here, one might conjecture that the possibility of excitation of
magnetic plasmon polaritons might provide a new avenue to the implementation of novel
techniques and devices
based on the interplay of photonics and magnetism at the nanoscale, similarly to the interplay
of photonics and electronics. Therefore, the implementation
of plasmonic chips for high data rate processing or efficient sensing applications might also be obtained based on
the excitation of magnetic plasmon polaritons. Finally, we have studied an ideal system in which
losses have been neglected, and future work including such effects is certainly forthcoming.

\begin{acknowledgments}
We are grateful to the Brazilian Agencies CNPq,
FAPESP, FAPERJ, and FUJB, the Colombian Agency
COLCIENCIAS, and CODI - Univ. of Antioquia for partial financial
support.
\end{acknowledgments}

\end{document}